\documentclass[%
 reprint,
 amsmath,amssymb,
 aps,
 pra,
]{revtex4-2}
\usepackage[utf8]{inputenc}
\usepackage[T1]{fontenc}
\usepackage{physics}
\usepackage{graphicx}
\usepackage{xcolor}
\usepackage{dcolumn}
\usepackage{bm}
\usepackage{appendix}
\usepackage{soul}
\usepackage{cancel}

\newcommand{\im}{\mathrm{i}}
\newcommand{\ex}{\mathrm{e}}
\newcommand{\J}{\mathrm{J}}

\newcommand{\de}{\mathrm{d}}

\begin{document}


\title{Light interactions with polar quantum systems}

\author{Piotr Gładysz}
\email{glad@umk.pl}
\author{Karolina Słowik}%
\email{karolina@fizyka.umk.pl} 
\affiliation{%
Institute of Physics, Faculty of Physics, Astronomy and Informatics, Nicolaus Copernicus University in Toruń, ul. Grudziądzka 5, 87-100 Toruń, Poland
}
%

\begin{abstract}
We investigate the dynamics of polar systems coupled to classical external beams in the ultrastrong coupling regime. The permanent dipole moments (PDMs) sustained by polar systems can couple to the electromagnetic field, giving rise to a variety of processes such as difference-frequency and harmonic generation, reflected in complicated dynamics of the atomic system. Here, we demonstrate that the atomic evolution can be described simply in a dynamic reference frame. We derive a Jaynes--Cummings-like framework with effective parameters describing frequency shift, coupling strength with the driving field, and a rescaled relaxation rate. The familiar linear scaling of the interaction strength with the field amplitude is replaced with a nonlinear dependence, suggesting potential applications for improving the coherence of quantum system ensembles with permanent dipoles.
\end{abstract}

\maketitle

\section{Introduction}
The interaction of light with matter is a fundamental phenomenon in physics, which underpins many modern problems in quantum optics and photonics \cite{dovzhenko2018,hertzog2019,rivera2020,smeets2023,marinho2024}. PDMs arise due to the lack of inversion symmetry in the variety of polar systems, such as giant dipole molecules \cite{kovarskii1999}, ultracold dipolar gases \cite{aymar2007}, quantum dots \cite{gawarecki2014}, Rydberg molecules \cite{booth2015, gonzalez2021}, and condensates thereof \cite{bigagli2024}. Unlike transition dipole moments that occur between pairs of system eigenstates, permanent dipoles are inherent properties of individual eigenstates and can significantly influence the system's response to external electromagnetic fields \cite{kibis2009, koppenhofer2016, chestnov2017, anton2017, gladysz2020, scala2021}.

Traditionally, the role of permanent dipoles in light-matter interaction has been considered minor, often relegated to causing trivial frequency shifts. However, recent studies have shown that these dipoles can induce coherent radiation at the Rabi frequency, leading to more complex and rich dynamical behaviors than previously understood \cite{kibis2009,meath2013,gladysz2020}. Polar systems have been investigated in the context of lasing at tunable frequencies \cite{chestnov2017}, radiation generation in tailored photonic environments \cite{savenko2012,izadshenas2023}, efficient two-photon excitation \cite{meath2013,meath2016}, light squeezing \cite{koppenhofer2016,anton2017}, Rydberg blockade \cite{guttridge2023}, ultracold chemistry \cite{moses2017, triana2021}, and quantum computations \cite{demille2002}. Materials with inherent $\chi^{(2)}$ nonlinearity were proposed for difference-frequency generation in a related mechanism \cite{barachati2015}. 
Similar effects were also studied in the magnetic context, where a rescaling of the Land\'{e} $g$ factor was found in atoms dressed with radiofrequency magnetic fields \cite{landre1970}.

In this manuscript, we investigate the dynamics of polar systems in the ultrastrong coupling regime with external beams, where the interaction strength between the electromagnetic field and the system is comparable to or exceeds the transition frequency of the system itself \cite{forn2019}. Despite the variety of processes that can occur in this regime, we reveal, through a series of unitary operations rotating the reference frame, that the system's behavior can be elegantly captured using a Jaynes--Cummings-type analytically solvable model.
This description, set in a dynamic reference frame, is based on three key parameters: the rescaled frequency shift, coupling strength, and relaxation rate. The rescaling is not merely a quantitative adjustment; it qualitatively transforms the behavior of the system: For example, the traditional linear dependence of interaction strength with the driving field amplitude is replaced by a nonlinear one, suggesting new potential regimes for the exploration of light-matter interactions.
We thoroughly verify the applicability range of the effective analytical description, confirming the robustness and accuracy of the model up to the ultrastrong coupling, where it shows good agreement with the full Hamiltonian dynamics but significantly reduces the computational cost.
Finally, based on the effective Hamiltonian, we find analytical expressions for resonance fluorescence spectra that would otherwise be challenging to access, due to the involvement of processes occurring at time scales that may differ by orders of magnitude, making the numerical integration challenging.

\section{Model}
\label{sec:model}

\begin{figure*}[ht]
    \begin{center}
        \includegraphics[scale=0.47]{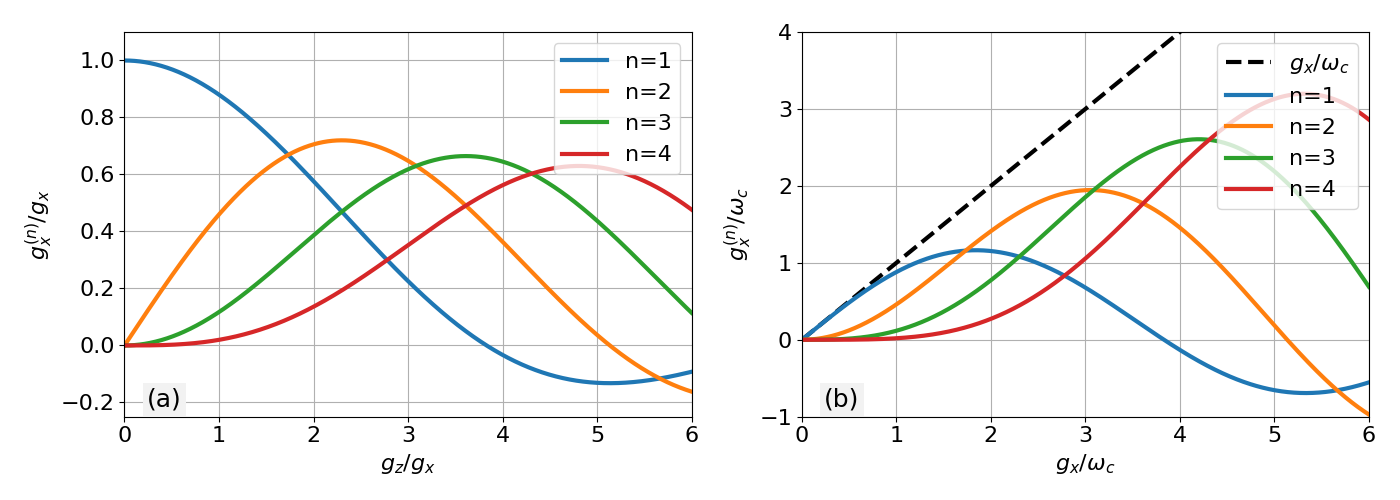}
    \end{center}
    \caption{
    Coupling strengths of first-order (blue), and higher-order (orange, green, red) resonances in two cases. (a) As functions of PDMs difference $g_z/g_x$ for a fixed value of the $\sigma_x$ coupling $g_x/\omega_c=1$. and relative to $g_x$. (b) As functions of the field's amplitude $g_x/\omega_c$ for a fixed value of PDMs difference $g_z/g_x=1$. The dashed line corresponds to the nonpolar system.
    }
\label{fig:fig1}
\end{figure*}

We consider a two-level polar quantum system (TLS) with the excited $\ket{e}$ and ground $\ket{g}$ states separated with the transition frequency $\omega_{eg}$. It is subject to a classical plane-wave laser drive $\vec{E}(t) = \vec{E}_0\cos(\omega_c t)$ with the amplitude $\vec{E}_0$ and frequency $\omega_c\approx\omega_{eg}$ near-resonant with the system transition. For a polar system, light couples with the full dipole moment $\vec{d}=\sum_{ij}\vec{d}_{ij}\ket{i}\bra{j}$ with $i,j\in\{e,g\}$, which includes the transition $\vec{d}_{i\neq j}$ and permanent $\vec{d}_{ii}$ elements. The Hamiltonian reads
\begin{equation}
    \begin{split}
    H(t)/\hbar &= \frac{1}{2}\omega_{eg}\sigma_z + \vec{E}(t)\cdot\vec{d} \\
    &= \frac{1}{2}\underbrace{\left(\omega_{eg} + g_z \cos(\omega_c t)\right)}_{\omega_{eg}(t)} \sigma_z 
    + g_x \cos(\omega_c t) \sigma_x
    \end{split}
\label{eq:full_Hamiltonian}
\end{equation}
Here, $\sigma_{x,z}$ are standard Pauli matrices. The drive induces transitions between the eigenstates with the transverse coupling strength \mbox{$g_x = \vec{E}_0\cdot\vec{d}_{eg} / \hbar$}, and a sinusoidal energy shift with the amplitude given by the longitudinal coupling strength $g_z = \vec{E}_0\cdot\Delta\vec{d}/\hbar$, where \mbox{$\Delta\vec{d}=\vec{d}_{ee}-\vec{d}_{gg}$}.
Although the transverse term $g_x$ also occurs in the standard nonpolar case, the longitudinal term $g_z$ is unique to systems with PDMs and is key to the effects investigated in this work. To quantify its impact, we introduce a parameter $\kappa_z = g_z/\omega_c$ which reduces to $0$ for nonpolar systems and is of the order of $1$ for polar systems in the longitudinal ultrastrong coupling regime, in which the dynamics is qualitatively altered. This should be distinguished from the "transverse" ultrastrong coupling regime defined by the relation $g_x/\omega_{eg}>1$.

A nonpolar system driven by light undergoes Rabi population oscillations with the Rabi frequency that for weak relaxation is given by $\Omega_\mathrm{R}\approx \sqrt{|g_x|^2+\delta^2}$, and depends on the drive amplitude and detuning $\delta=\omega_c-\omega_{eg}$ \cite{loudon}. In a polar system, the eigenstate energies oscillate inducing a time-dependent detuning according to Eq.~(\ref{eq:full_Hamiltonian}) $\delta(t)=\delta+g_z\cos\omega_c t$, leading to complicated dynamics mixing a range of oscillation frequencies. In the ultrastrong coupling regime, these contributions should not be neglected \textit{a priori} through a rotating-wave-type approximation. Here, we document in detail a transition to a reference frame that allows one to capture this complicated behavior in an analytically solvable model. The transition involves a series of noncommutative reference frame rotations realized by unitary operators $U$ that transform the Hamiltonian as $H\rightarrow UHU^\dagger + \im \hbar \dot{U} U^\dagger$, where the dot stands for time derivative, and dagger $^\dagger$ indicates Hermitian conjugate. These rotations subsequently simplify the form of the Hamiltonian.

\subsection{Transformation $U_1$ -- revealing higher-order resonances}
\label{sec:U1}
With the first transformation given by Kibis in \cite{kibis2009} $U_1 = \exp({\frac{1}{2}\im\kappa_z\sin(\omega_c t)\sigma_z})$, we remove the time dependence of the term proportional to the $\sigma_z$ operator at the cost of a more complex $\sigma_x$ coupling split into time-dependent sum of flip operators $\sigma^\pm$
\begin{equation}
    \begin{split}
    &H_1/\hbar = \frac{1}{2}\omega_{eg} \sigma_z\\
    &+ g_x\cos(\omega_c t)(\ex^{\im\kappa_z\sin(\omega_c t)}\sigma^+ + \ex^{-\im\kappa_z\sin(\omega_c t)}\sigma^-).
    \label{eq:Hamiltonian1raw}
    \end{split}
\end{equation}
Using the Jacoby-Anger identity $\ex^{\pm \im \kappa_z \sin(\omega_c t)} = \sum^{\infty}_{n=-\infty} \J_n(\kappa_z)\ex^{\pm\im n\omega_c t}$ which involves Bessel functions of the first kind $\J_n$, we can rewrite Eq.~\eqref{eq:Hamiltonian1raw} as 
\begin{widetext}
\begin{equation}
    H_1/\hbar = 
    \frac{1}{2} \omega_{eg} \sigma_z
    +\frac{1}{2}\sum_{n=1}^\infty \underbrace{\frac{2n}{\kappa_z}\J_n(\kappa_z)g_x}_{g_x^{(n)}}\left(\ex^{\im n\omega_c t} + [-1]^{n+1} \ex^{-\im n\omega_c t}\right)
    \left(\sigma^++[-1]^{n+1}\sigma^-\right).
\label{eq:Hamiltonian1separated}
\end{equation}
\end{widetext}
With that form, we separate terms by the order of the resonance $n\omega_c$, where terms proportional to $g_z$ are incorporated in the effective coupling strengths $g_x^{(n)}$. Bunching together terms with $n\geq2$ and naming them $H_{ho}/\hbar$, we end up with an effective Rabi-like model
\begin{equation}
    H_1/\hbar = \frac{1}{2}\omega_{eg}\sigma_z + g_x^{(1)}\sigma_x\cos\omega_c t + H_{ho}/\hbar,
\label{eq:Hamiltonian1}
\end{equation}
where $H_{ho}$ contains all the higher-order terms. It is worth noticing that the polarity of the system "unlocks" even-order transitions ($n=2, 4, 6, \dots$) forbidden in inversion-symmetric systems.

This step involves no approximations as Eqs.~\eqref{eq:Hamiltonian1separated} and~\eqref{eq:Hamiltonian1} are exact -- the polar terms are fully included. In a frame rotating with a sinusoidally varying frequency, the system behaves as if it were not polar but coupled with an electromagnetic field through a modified coupling strength $g_x^{(1)}$, nonlinearly dependent on the field amplitude. As we verify later in this paper, the corrections arising from $H_{ho}$ may become important for the case of the ultrastrong $\sigma_x$ coupling.

For the fixed coupling strength $g_x/\omega_c = 1$, the values of the couplings as functions of the polarity $g_z/g_x$ of the system are presented in Fig.~\ref{fig:fig1}a. For small PDMs and field amplitudes -- typical in optical experiments and calculations -- the first-order coupling $g_x^{(1)}$ is dominant. However, highly polarized systems coupled to strong fields may require accounting for higher-order transitions through terms with $n\geq2$. Most importantly, as shown in Fig.~\ref{fig:fig1}b, for the ratio $g_z/g_x=1$, the coupling strength $g_x^{(1)}$ significantly deviates from standard $g_x$ (dashed line) as we increase the driving field's amplitude $g_x/\omega_c$. This nonlinear behavior is investigated further in the next parts.

\subsection{Transformation $U_2$ -- dealing with counter-rotating terms}
\label{sec:U2}
In the following step, counter-rotating terms are similarly incorporated in effective parameters through the transformation $U_2 = \exp({\im\xi\kappa_x\sin(\omega_c t)\sigma_x})$ given by Yan \cite{yan2013}, where $\kappa_x=g_x^{(1)}/\omega_c$. In result, full Hamiltonian \eqref{eq:Hamiltonian1} transforms into a rather complex form consists of terms with different oscillation frequencies ($\omega_c$, $2\omega_c$, $3\omega_c$, \dots):
\begin{widetext}
\begin{equation}
    \begin{split}
    &H_2/\hbar = \\
    &\frac{1}{2} \omega_{eg} \left[ \left( \J_0(2\xi\kappa_x) + \sum\limits^\infty_{k=1} \J_{2k}(2\xi\kappa_x) \left( 
    \ex^{\im 2k\omega_c t} + \ex^{-\im 2k\omega_c t} \right) \right) \sigma_z - \im \sum\limits^\infty_{k=0} \J_{2k+1}(2\xi\kappa_x) \left( 
    \ex^{\im (2k+1)\omega_c t} - \ex^{-\im (2k+1)\omega_c t} \right) \sigma_y \right]\\
    &+\frac{1}{2} \sum\limits^\infty_{k=0}g^{(2k+1)}_x(\kappa_z) \left( 
    \ex^{\im (2k+1)\omega_c t} + \ex^{-\im (2k+1)\omega_c t} \right) \sigma_x - \xi g^{(1)}_x \cos(\omega_c t) \sigma_x\\
    &+\frac{\im}{2} \sum\limits^\infty_{k=1}g^{(2k)}_x(\kappa_z) \left( 
    \ex^{\im 2k\omega_c t} - \ex^{-\im 2k\omega_c t} \right) \times\\
    &\left[ \left( \J_0(2\xi\kappa_x) + \sum\limits^\infty_{l=1} \J_{2l}(2\xi\kappa_x) \left( 
    \ex^{\im 2l\omega_c t} + \ex^{-\im 2l\omega_c t} \right) \right) \sigma_y + \im \sum\limits^\infty_{l=0}\J_{2l+1}(2\xi\kappa_x) \left( 
    \ex^{\im (2l+1)\omega_c t} - \ex^{-\im (2l+1)\omega_c t} \right) \sigma_z \right].
    \end{split}
\label{eq:Hamiltonian2separated}
\end{equation}    
\end{widetext}
Following the idea provided in [30] for systems without PDMs, we introduced the parameter $\xi\in(0,1)$ which fulfills the equation 
\begin{equation*}
    \omega_{eg}\J_1(2\xi\kappa_x) = (1-\xi)g^{(1)}_x \equiv \frac{1}{2}\Omega,
\label{eq:xi_evaluation}
\end{equation*}
where $\Omega$ plays the role of an effective coupling strength. It is now possible to separate the different frequency terms in the spirit of the RWA. Keeping time-independent terms and the ones oscillating with the frequency $\omega_c$, we arrive at an approximation of the Hamiltonian from Eq.~\eqref{eq:Hamiltonian2separated} relevant for the resonance fluorescence problem
\begin{widetext}
\begin{equation}
    H_2/\hbar \approx \frac{1}{2}\underbrace{\omega_{eg}\J_0(2\xi\kappa_x)}_{\omega_{eg}^\prime } \sigma_z+\underbrace{\sum_{k=1}^\infty g_x^{(2k)}(\kappa_z)\frac{4k}{2\xi\kappa_x}\J_{2k}(2\xi\kappa_x)}_{ g_z^\prime }\cos(\omega_c t)\sigma_z
    + \frac{1}{2}\Omega\left(\ex^{-\im\omega_c t}\sigma^+ + \ex^{\im\omega_c t}\sigma^- \right).
\label{eq:Hamiltonian2}
\end{equation}
\end{widetext}

\begin{figure*}[ht]
    \begin{center}
    \includegraphics[scale=0.47]{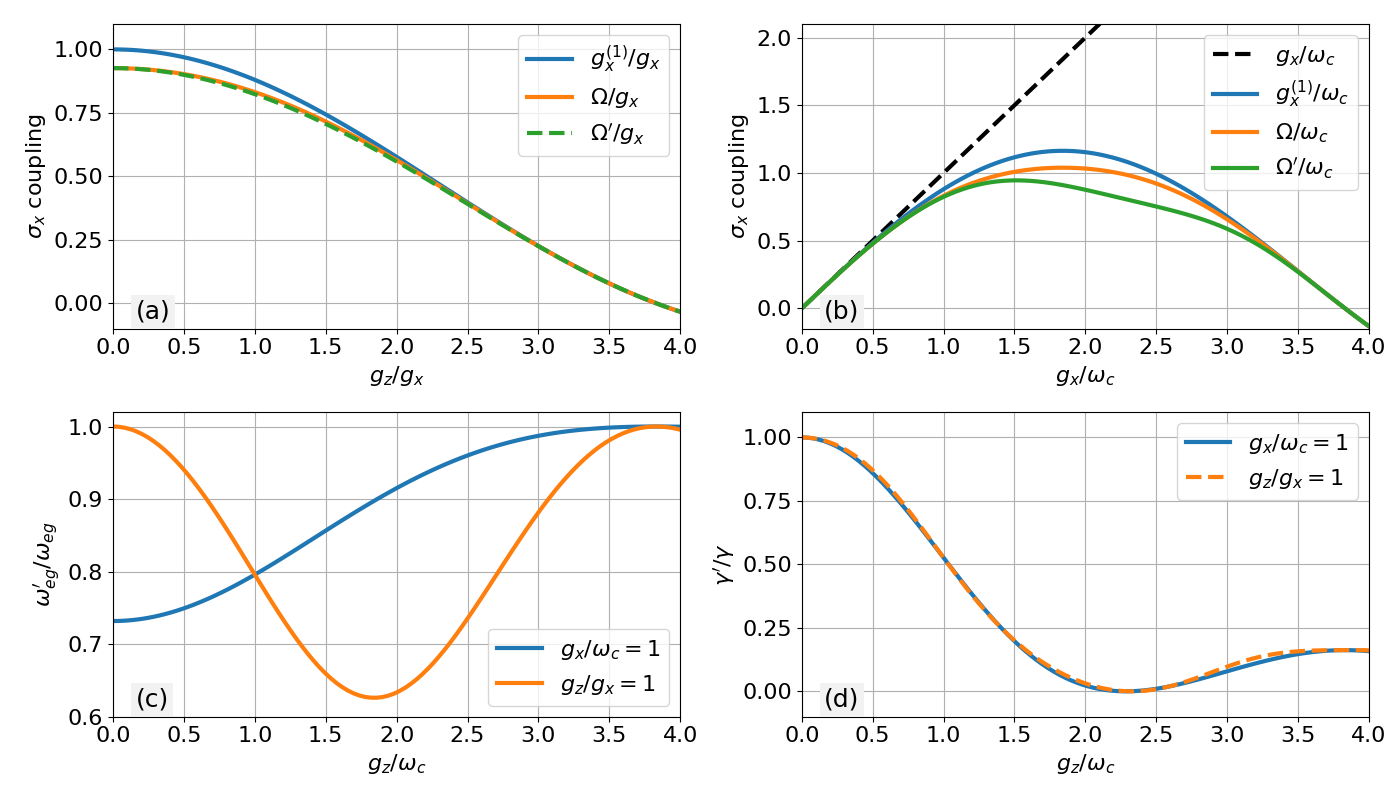}
    \end{center}
    \caption{
    The upper plots show comparisons of the $\hat{\sigma}_x$ coupling strengths for different corrections as a function of 
    (a) PDMs difference (for $g_x/\omega_c=1$), (b) driving field amplitude (for $g_z/\omega_c=1$). The lower plots present: (c) the normalized effective transition frequency, (d) the normalized effective gamma rate, as functions of $g_z$ coupling, both in cases of fixed driving field (blue) or PDMs difference (orange), respectively.
    }
\label{fig:fig2}
\end{figure*}
\noindent Here, we introduced $\omega_{eg}^\prime $, which is the modified resonance frequency shifted due to the influence of the counter-rotating terms. Additionally, the coupling $g_z^\prime$ is a correction from the counter-rotating terms to the polar coupling of the system. In summary, with the second transformation, we have taken into account the counter-rotating terms in the effective parameters $\omega_{eg}^\prime$, $g_z^\prime$, and $\Omega$.

\subsection{Transformation $U_3$ -- effectively removing the $\sigma_z$ coupling}
\label{sec:U3}
The form of the Hamiltonian given by Eq.~\eqref{eq:Hamiltonian2} is very similar to the initial one from Eq.~\eqref{eq:full_Hamiltonian}. Hence, the oscillating $\sigma_z$ term can be treated with a transformation analogous to $U_1$, namely $U_3 = \exp({\frac{1}{2}\im\kappa^\prime_z\sin(\omega_c t)\sigma_z})$, where $\kappa_z^\prime = g_z^\prime/\omega_c$. The resulting Hamiltonian has the Jaynes--Cummings form as it occurs in the Schr\"{o}dinger-picture
\begin{equation}
    H_3/\hbar \approx \frac{1}{2}\omega_{eg}^\prime  \sigma_z 
    + \frac{1}{2}\underbrace{\Omega\J_0(\kappa^\prime_z)}_{{\Omega^\prime}}\left(\ex^{-\im\omega_c t}\sigma^+ + \ex^{\im\omega_c t}\sigma^- \right),
\label{eq:Hamiltonian3}
\end{equation}
with $\Omega^\prime$ being slightly modified coupling strength due to the $\kappa_z^\prime$ occurrence. The terms containing higher-order Bessel functions $\J_n(\kappa_z^\prime)$ for $n\geq1$ are neglected due to the fact that $\kappa_z^\prime \ll 1$.

\subsection{Transformation $U_4$ -- moving to the interaction picture}
\label{sec:U4}
Finally, the transformation $ U_4 = \exp({\frac{1}{2}\im\omega_c t\sigma_z})$ brings Hamiltonian in Eq.~\eqref{eq:Hamiltonian3} to the interaction picture, where it is time-independent and acquires the analytically solvable form
\begin{equation}
    H_\mathrm{eff}/\hbar = -\frac{1}{2}\underbrace{\left( 
\omega_c-\omega_{eg}^\prime \right)}_{\delta^\prime}\sigma_z + \frac{1}{2}\Omega^\prime\sigma_x,
\label{eq:HamiltonianEff}
\end{equation}
where $\delta^\prime$ is the effective detuning from the resonance. On the way to derive the effective Hamiltonian, in Eq.~\eqref{eq:Hamiltonian2}, we have omitted higher-order corrections negligible in the regime investigated below. 

The Hamiltonian given by Eq.~(\ref{eq:HamiltonianEff}) is analytically solvable, and yet, accounts for a plethora of physical effects related to the system's asymmetry \textit{via} PDMs, transverse and longitudinal ultrastrong couplings, Bloch--Siegert shift, and beyond, and is exact up to higher-order effects. The rich variety of the captured phenomena is reflected in the effective parameters that we now discuss.

The effective coupling strength has a complicated functional dependence on the field amplitude and atomic dipole moments
\begin{equation*}
    \Omega^\prime = \J_0(\kappa^\prime_z)\Omega \approx g_x^{(1)},
\label{eq:Omega_prime}
\end{equation*}
where the latter approximate expression is valid for moderately strong interactions with $\vec{E}_0\Delta\vec{d}< \hbar\omega_c$.
Therefore, we have found an iterative sequence of coupling strengths $g_x^{(1)}\rightarrow \Omega \rightarrow \Omega^\prime$ obtained through subsequent rotations.
The effective coupling strengths are plotted in Fig.~\ref{fig:fig2}(a,b) respectively as a function of the PDMs difference in the TLS eigenstates and of the electric field amplitude.
The blue, orange, and green lines represent coupling strengths after subsequent frame rotations, and the black dashed line is the usual coupling strength linearly scaled with the field amplitude in the original reference frame. 

The field detuning $\delta^\prime$ is defined with respect to the effective transition frequency $\omega_{eg}^\prime$, whose dominant Taylor expansion coefficient
\begin{equation*}
    \xi^2 \omega_{eg}\frac{J_1(\kappa_z)^2|2g_x|^2}{\kappa_z^2\omega_c^2}\xrightarrow[\kappa_z\rightarrow 0]{}\xi^2\omega_{eg}\frac{|g_x|^2}{\omega_c^2}
\end{equation*}
can be recognized as the Bloch--Siegert shift \cite{bloch1940}. The scaling of the effective frequency $\omega_{eg}^\prime$ with normalized $g_z$ coupling is shown in Fig.~\ref{fig:fig2}(c).

To conclude, we have applied a series of noncommutative unitary transformations or Bloch sphere rotations around the $z$ and $x$ axes. Each of these transformations introduces a Hamiltonian simplification and captures subsequent Hamiltonian terms in effective parameters. As a result, we have arrived at the Hamiltonian $H_\text{eff}$ given by Eq.~\eqref{eq:HamiltonianEff}, whose simple form captures the rich physics of interactions of light with transition and permanent dipole moments. In $H_\text{eff}$, we have neglected higher-order resonances of the type $n\omega_c\approx\omega_{eg}$, as is justified with our original assumption that $\omega_c\approx\omega_{eg}$.

\subsection{Dissipative dynamics}
\label{sec:Dissipation}
To account for dissipative dynamics, we have augmented the Hamiltonian Eq.~(\ref{eq:full_Hamiltonian}) with a coupling to a bosonic thermal bath of the form
\begin{equation}
    H^R/\hbar = \sum_p \omega_p b^\dagger_p b_p + \frac{1}{2}\sum_p g_p (b^\dagger_p + b_p)\sigma_x,
\label{eq:HamiltonianR}
\end{equation}
where reservoir modes are described by the creation $b_p^\dagger$ and annihilation $b_p$ operators with corresponding mode frequencies $\omega_p$. The coupling strength between these modes and the system is given by $g_p$. We move to the interaction picture for the reservoir by making use of the unitary transformation $U^R_\text{int} = \exp({\im\sum_m \omega_m b^\dagger_m b_m t})$ that does not affect the system and, therefore, does not interfere with our previous calculations. The Hamiltonian form Eq.~\eqref{eq:HamiltonianR} in the interaction picture takes the form
\begin{equation}
    H^R_\text{int}/\hbar = \frac{1}{2}\sum_p g_p \left(b^\dagger_p\ex^{\im\omega_p t} + b_p\ex^{-\im\omega_p t}\right)\sigma_x.    
\label{eq:HamiltonianRint}
\end{equation}

Next, we perform the same reference frame rotations ($U_1$ -- $U_4$), and consider a thermal reservoir in the vacuum state weakly coupled to the system, which leads to the approximated form of the system-reservoir interaction Hamiltonian
\begin{equation}
    H^R_\text{int}/\hbar \approx
    \frac{1}{2}\sum_p g_p^\prime \left(b^\dagger_p\sigma^-\ex^{\im(\omega_p - \omega_c)t} + b_p\sigma^+\ex^{-\im(\omega_p - \omega_c)t}\right),
\label{eq:HamiltonianRintApprox}
\end{equation}
where $g_p^\prime\approx g_p\J_0(\kappa_z)\J_0(\kappa^\prime_z)$ is an effective coupling strength to the $p$-mode of the reservoir. The full transformed Hamiltonian from Eq.~\eqref{eq:HamiltonianRint}, non-approximated form of the $g_p^\prime$ parameter from Eq.~\eqref{eq:HamiltonianRintApprox}, and the intermediate unitary transformations, are provided in Appendix~\ref{sec:AppendixA}.

As a result, following the Weisskopf-Wigner theory with the Markovian approximation, we find a rescaled spontaneous emission rate which for $\vec{E}_0\Delta\vec{d}\leq \hbar\omega_c$ takes the form
\begin{equation}
    \gamma^\prime = \left(\frac{g_p^\prime}{g_p}\right)^2 \gamma \approx \J_0(\kappa_z)^2\J_0(\kappa^\prime_z)^2\gamma,
\label{eq:gamma_prime}
\end{equation}
and is further modified for strong fields (Appendix~\ref{sec:AppendixA}).
Thus, in the dynamic frame, the spontaneous emission rate in polar systems depends on the external field amplitude $E_0$ [see Fig.~\ref{fig:fig2}(d), where the rate is shown without approximations]. The emission rate keeps its original value $\gamma$ for vanishing PDMs or in the absence of an external field. 

The simple form of the effective Hamiltonian given by Eq.~(\ref{eq:HamiltonianEff}) along with the effective emission rate from Eq.~\eqref{eq:gamma_prime} allows us to analytically evaluate the stationary expectation values of the induced dipole moment and the population inversion in the rotated frame
\begin{subequations}
    \begin{align}
    &\langle\sigma^\pm\rangle_s = \frac{\Omega^\prime(2\delta^\prime\pm\im\gamma^\prime)}{2{\Omega^\prime}^2+4{\delta^\prime}^2+{\gamma^\prime}^2},\\
    &\langle\sigma_z\rangle_s = -\frac{(4{\delta^\prime}^2+{\gamma^\prime}^2)}{2{\Omega^\prime}^2+4{\delta^\prime}^2+{\gamma^\prime}^2}.
    \end{align}
\label{eq:stationarysolutions}
\end{subequations}
Details are provided in Appendix~\ref{sec:AppendixB}.

\section{Fluorescence spectra}
\label{sec:spectra}
Building upon stationary solutions, we use the Onsager-Lax theorem to evaluate the fluorescence spectrum of the system following the approach proposed in \cite{yan2013} for nonpolar systems. All the calculations are provided in details in Appendix~\ref{sec:AppendixC}. At this step, the price for the simplicity of the Hamiltonian is the complicated form of the evolution operator in the rotating frame that makes the resulting expressions quite complex. Yet, we obtain their analytical forms predicting a sequence of Mollow triplets centered at multiples $n\omega_{c}$ of the illumination frequency, with sidebands separated by the effective Rabi frequency $\widetilde{\Omega}$, which is calculated based on the coupling strength $\Omega^\prime$ and $\gamma^\prime$-rate such that $\widetilde{\Omega}\xrightarrow{} \Omega^\prime$ for $\gamma^\prime \xrightarrow{} \gamma$. For the full derivation of the lineshapes and $\widetilde{\Omega}$ parameter see Appendix~\ref{sec:AppendixC}.

\begin{figure}[t]
    \begin{center}
        \includegraphics[width=\columnwidth]{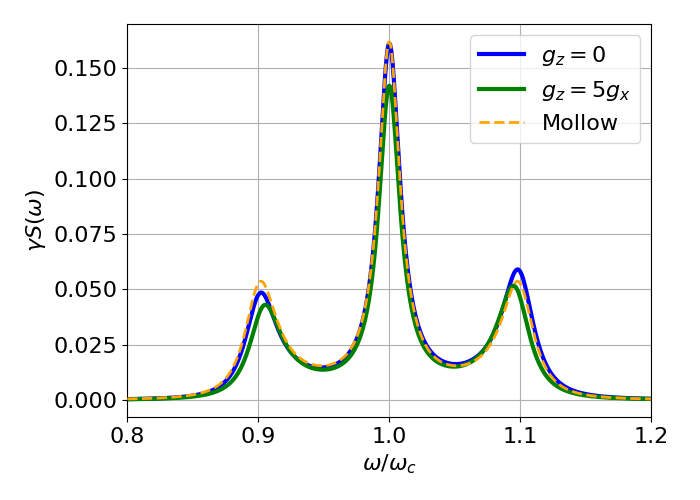}
    \end{center}
    \caption{
    Comparison of spectra calculated for the two-level system with (green line), and without (blue line) PDMs, and the Mollow-triplet (dashed line), calculated for the resonant case $\omega_c = \omega_{eg}$. The coupling strength $g_x/\omega_c = 0.1$ and the relaxation rates $\gamma/\omega_c = \gamma^\prime/\omega_c = 0.02$ were assumed to be the same for all cases. In the polar case, $g_z = 5g_x$.
    }
\label{fig:fig3}
\end{figure}

\begin{figure*}[ht]
    \begin{center}
        \includegraphics[scale=0.5]{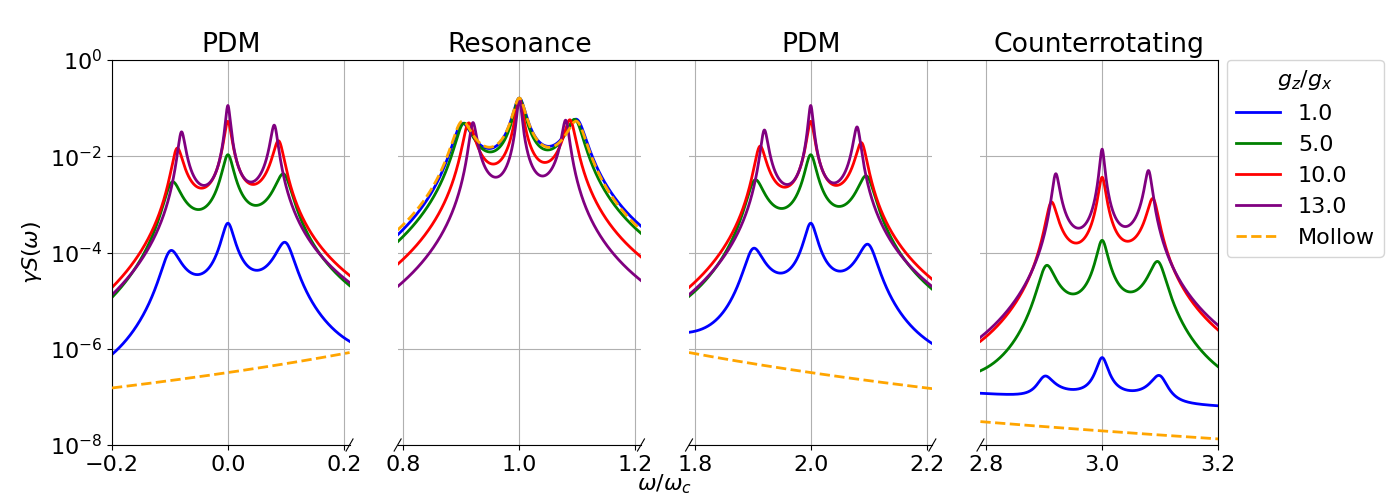}
    \end{center}
    \caption{
    Comparison of spectra for several different values of the PDMs ($g_z/g_x$ ratios) and compared to the results obtained by Mollow (dashed line), obtained for the resonant case $\omega_c = \omega_{eg}$, $\gamma/\omega_c=0.02$, $g_x/\omega_{c} = 0.1$. In the plot, calculated $\gamma^\prime$ values are used. The titles over respective parts correspond to the origin of the triplet: "PDM" indicates the existence of the peaks due to the polarity of the system, while "Resonance" and "Counterrotating" parts are typical for any electric-dipole interaction with a strong, resonant, external electromagnetic field. 
    }
\label{fig:fig4}
\end{figure*}

\begin{figure*}[ht]
    \begin{center}
        \includegraphics[scale=0.5]{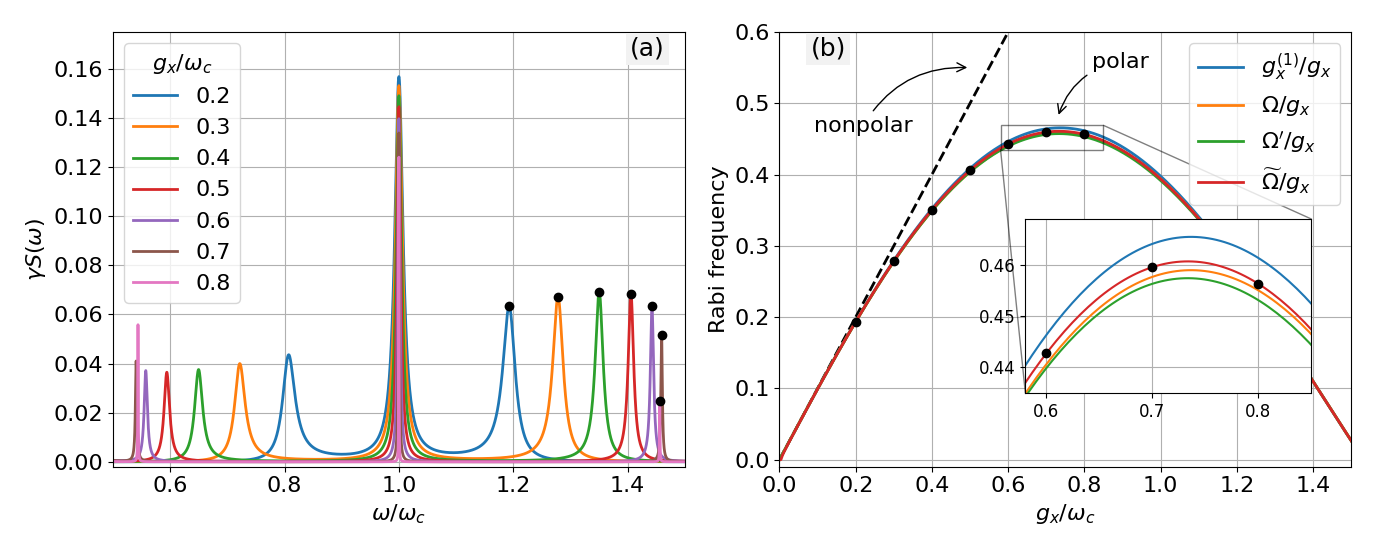}
    \end{center}
    \caption{
    (a) Resonance fluorescence spectra for different values of the $\hat{\sigma}_x$ coupling $g_x$ and fixed PDMs difference $g_z/g_x = 2.5$. For each spectrum $\omega_c = \omega_{eg}$, and spontaneous emission rates $\gamma^\prime$s are calculated based on $\gamma/\omega_c = 0.02$. 
    (b) Position of the peaks of the right sidebands shown on top of the curves with approximations to the evaluation of the Rabi frequency. The dashed line is shown as a reference for the nonpolar case.
    }
\label{fig:fig5}
\end{figure*}
Let us first analyze the resulting spectrum of a polar system illuminated at $\omega_c = \omega_{eg}$ around the original peak, i.e., in the vicinity of the illumination frequency. This result is compared with the one obtained for the nonpolar case \cite{yan2013} and the analytical results derived by Mollow \cite{mollow1969} (Fig.~\ref{fig:fig3}).
The relatively strong field, and hence the significant impact of the counter-rotating terms is responsible for the asymmetry of the sidebands in both polar and nonpolar cases but is not captured by the Mollow solution. This asymmetry is comparable in the two former cases, hence we conclude that the existence of the PDMs does not influence the counter-rotating terms. However, it provides a significant resonance shift and correction to the transverse coupling strength, which is crucial for the evaluation of the peak amplitudes and spectral positions of the sidebands, given by the effective coupling~$\widetilde{\Omega}$.

We now proceed to discuss the spectral features centered at $n\omega_{c}$. Neglecting higher-order corrections, the relevant terms correspond to $n=0,1,2,3$, as predicted in earlier studies \cite{scala2021}.
In systems without PDMs, the emission at even multiples of the laser frequency does not occur \cite{meath2013}. The peak at $3\omega_{eg}$ is induced in this case by the counter-rotating terms in the Rabi Hamiltonian \cite{braak2011,scala2021}.
Polarity in the system reduces the Hamiltonian symmetry, gives rise to the even-order fluorescence peaks, and modifies the odd ones. The low-energy emission at Rabi frequency $\widetilde{\Omega}$ discussed previously in various contexts \cite{kibis2009,chestnov2017,gladysz2020}, is captured here as the Mollow sideband of the triplet corresponding to $n=0$.
In Fig.~\ref{fig:fig4}, we demonstrate the spectra for $g_x/\omega_c = 0.1$ and $g_z/\omega_c = 0.1$, $0.5$, $1.0$ and $1.3$. We assume that the laser is tuned to the original atomic resonance $\omega_c = \omega_{eg}$. Again, the field gives rise to a large frequency shift, hence, a significant effective detuning responsible for the asymmetry of the sidebands.
With increasing PDMs difference, the light interaction with the atomic system becomes dominated by related terms, in the original Hamiltonian proportional to $g_z$. This is reflected in the emission spectrum through the modified relative intensities of the peaks: The original Mollow triplet at $\omega_c$, dominant for weakly polar systems with small PDMs, is gradually suppressed as the emission is directed to the other frequency channels.

Finally, Fig.~\ref{fig:fig5}(a) shows how the spectral position of the sidebands can be tuned with the driving field amplitude. Here, we assume again the illumination at $\omega_c = \omega_{eg}$. The ratio of permanent- and transition-dipole moment element components parallel to the field is equivalent to the ratio of coupling constants, that we fix at \mbox{$g_z/g_x = 2.5$}. The relaxation rate is evaluated according to Eq.~(\ref{eq:gamma_prime}) with initial value $\gamma/\omega_c=0.02$.
We find that as the field amplitude increases, the sidebands are initially pulled away from the central peak at a decreasing pace and eventually turn back towards the center at $\omega_c$. Additionally, by increasing the field amplitude we decrease both the transition frequency $\omega^\prime$ and spontaneous emission $\gamma^\prime$-rates according to the behavior shown in Fig.~\ref{fig:fig2}(c,d). As a result, we observe the narrowing and growth (shrinking) of the right (left) sidebands.

In the Jaynes--Cummings model, the difference in the spectral position of the sideband and the central peak corresponds to the coupling strength. Naturally, the same is expected from the effective description that we have reduced to the Jaynes--Cummings form. Indeed, as can be seen in Fig.~\ref{fig:fig5}(b), the position of spectral sidebands given by $\widetilde{\Omega}$ is in good agreement with the prediction of the effective coupling strength $\Omega^\prime$. For comparison, we also show the coupling strengths' predictions evaluated after subsequent transformations $U_1$ -- $U_4$. As it is clear from the inset, all frame rotations need to be performed to eventually obtain the excellent agreement of $\widetilde{\Omega}$ (red line) with the calculated spectral positions of sidebands (black dots). However, already the first transformation provides a qualitatively correct expression for the effective coupling strength $g_x^{(1)}$, as can be concluded from the examination of the blue line.

\begin{figure*}[ht]
    \begin{center}
        \includegraphics[width=\linewidth]{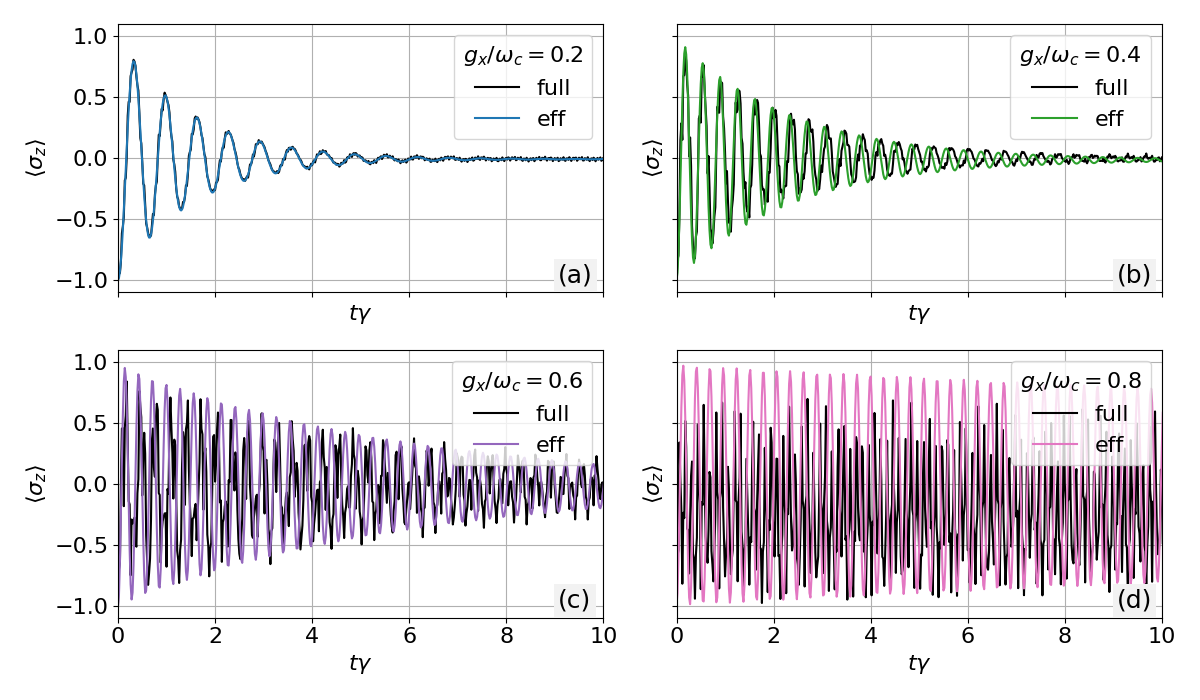}
    \end{center}
    \caption{
    Evolution of the population represented by $\langle \hat{\sigma}_z \rangle$. Each plot corresponds to one of the spectral lineshapes presented in Fig.~\ref{fig:fig5}a (color matching),
    }
    \label{fig:fig6}
\end{figure*}

\section{Applicability and real systems}
\label{sec:applicability}
We aim to comment on the applicability of the derived model in Fig.~\ref{fig:fig6} presenting the dynamics of the population in the system for several external field amplitudes. 
The numerical calculations were made by solving the master equation with the full Hamiltonian \eqref{eq:full_Hamiltonian} and taking into account the spontaneous emission process with the rate $\gamma=0.02\omega_c$
\begin{equation}
\begin{split}
    \dot{\rho}(t) &= [H(t)/\hbar, \rho]\\
    &- \im\frac{\gamma}{2}\left(\sigma^+\sigma^-\rho(t)+ \rho(t) \sigma^+\sigma^- - 2\sigma^-\rho(t) \sigma^+ \right),
\end{split}
\label{eq:master}
\end{equation}
where $\rho$ denotes the density operator of the polar system.
This result is compared with the one obtained with the derived Hamiltonian $H_\mathrm{eff}$ and the modified decay rate $\gamma^\prime$.

Each panel corresponds to a selected spectrum from Fig.~\ref{fig:fig5}a for a fixed ratio $g_z/g_x=2.5$ (matching colors are used). Note that for $g_x/\omega_c=0.4$, longitudinal ultrastrong $\sigma_z$ coupling is already reached with $g_z=\omega_c$. Despite the simplicity of the effective Hamiltonian described with a pair of effective parameters, we obtain very good agreement with the full Hamiltonian dynamics even near the Bessel function peak. We can thus capture the complex dynamics of systems with permanent dipoles subjected to extremely strong electromagnetic fields with this simple model. 
Beyond $g_x/\omega_c=0.6$ [Fig.~\ref{fig:fig6}(c)], higher-order corrections begin to play a more important role so that the effective dynamics ceases to predict the qualitative behaviour of the system. Moreover, in this regime the suppression of the $\gamma^\prime$-rate is significant and in Fig.~\ref{fig:fig6}(c,d) leads to the reduction of the oscillations' damping. The higher-order corrections could also alter this behavior.

On the other hand, in our study we focused on the systems with significantly larger $\sigma_z$ coupling strength compared to the $\sigma_x$ one, e.g., the data plotted in Fig.~\ref{fig:fig4} with the ratio up to $g_z/g_x=13$. With that, the observation of the phenomena originating from the existence of the PDMs are pronounced, while driving field intensity is still in the reasonable regime. The information of the PDM values in the excited electronic states ($d_{ee}$) is limited in the literature as the measurement of such quantity is challenging, hence, for the sake of estimation, we focus on the available data for the ground states, assuming $\Delta \vec{d} \approx \vec{d}_{gg}$. Additionally, we neglect the vector character of the dipole moments and estimate $g_z/g_x\approx\Delta d/d_{eg}$. 

Real-life examples include small molecules and asymmetric quantum dots as they naturally exhibit high permanent-to-transition dipole ratios in free space. For instance, the results obtained with the Turbomole software~\cite{furche2014,balasubramani2020} for diatomic molecules provide the following ratios: $d_{gg}/d_{eg} \approx 11.1$ for KBr, $d_{gg}/d_{eg} \approx 60.8$ for BaF, and $d_{gg}/d_{eg} \approx 19.5$ for SrO. Extensive datasets for small molecules can be found in literature \cite{aymar2007, quemener2012, gadway2016, liu2020}. Moreover, analyzing transitions between rotational levels instead of the electronic ones enables further control of these ratios, as we have shown in studies of LiH. There, the permanent and transition dipoles can be tuned as they orient in external static field \cite{gladysz2020}. In asymmetric quantum dots, $g_z/g_x$ reflects the spatial asymmetry induced by structural or field-induced polarization, which creates a permanent dipole \textit{via} spatial separation of the exciton components (electron and hole). While this asymmetry enhances the permanent dipole moment, it may reduce the transition dipole due to diminished overlap between the electron and hole wavefunctions of the exciton. This interplay underscores the tunability of $g_z/g_x$ in quantum dots, governed by geometric design and external field strength. The permanent dipole moments may be of the order of tens of Debyes in InGaN and InAs on GaAs quantum dots \cite{schulhauser2002, ostapenko2010}.

\section{Conclusions}
\label{sec:conclusions}
In conclusion, we explored the behavior of polar systems under transverse and longitudinal ultrastrong couplings with classical external beams. By switching to a dynamic reference frame using a series of unitary operations, we derived an analytically solvable Jaynes--Cummings-like model. This model captures the impact of PDMs, counter-rotating terms, and strong fields in effective frequency shift, coupling strengths, and relaxation rate, helping us understand the complex interactions in these systems.

Our findings show that permanent dipoles play a crucial role in changing how the system interacts with external fields, leading to a nonlinear interaction strength. This shift from traditional linear scaling suggests new avenues for research on light interactions with polar matter, especially in enhancing the coherence of multiple quantum systems with permanent dipoles. Future research should investigate the identified nonlinear effects and their practical applications, explore more complex quantum systems, and study their behavior under different external conditions. Additionally, expanding this framework to include interactions at higher-order resonances and multipartite dynamics could offer further insights into the principles that govern this interesting interaction regime.

\section*{Acknowledgements}
The authors thank Florian Weigend from the Karlsruhe Institute of Technology, Germany, for providing results obtained using the Turbomole software~\cite{furche2014}, as referenced in Section~\ref{sec:applicability}. 

This work was supported by the National Science Centre, Poland, under the PRELUDIUM program (grant number 2021/41/N/ST2/02068).

\begin{appendices}
\begin{widetext}
\section{Appendix: Derivation of relaxation rate equation}
\label{sec:AppendixA}
To take into account dissipation processes (e.g., spontaneous emission), we consider an interaction between the TLS and a bosonic reservoir given by Hamiltonian Eq.~\eqref{eq:HamiltonianR} further transformed to the interaction picture in Eq.~\eqref{eq:HamiltonianRint}.

We now apply all the transformations $U_1$--$U_4$ to the system operators $\sigma_{x,y,z}$, and list below the nontrivial transformation prescriptions:
\begin{equation*}
\begin{split}
    &U_1\sigma_xU^\dagger_1 =
    \left( \J_0(\kappa_z) + \sum_{k=1}^{\infty}\J_{2k}(\kappa_z)\left(\ex^{\im 2k \omega_c t} + \ex^{-\im 2k \omega_c t}\right)\right)\sigma_x
    +\sum_{k=0}^{\infty}\J_{2k+1}(\kappa_z)\left(\ex^{\im (2k+1) \omega_c t} - \ex^{-\im (2k+1) \omega_c t}\right)\im\sigma_y,\\
    &U_2\im\sigma_yU^\dagger_2 =
    \left( \J_0(2\xi\kappa_x) + \sum_{k=1}^{\infty}\J_{2k}(2\xi\kappa_x)\left(\ex^{\im 2k \omega_c t} + \ex^{-\im 2k \omega_c t}\right)\right)\im\sigma_y
    -\sum_{k=0}^{\infty}\J_{2k+1}(2\xi\kappa_x)\left(\ex^{\im (2k+1) \omega_c t} - \ex^{-\im (2k+1) \omega_c t}\right)\sigma_z,\\
    &U_3\sigma_1U^\dagger_3 =
    \left( \J_0(\kappa^\prime_z) + \sum_{k=1}^{\infty}\J_{2k}(\kappa^\prime_z)\left(\ex^{\im 2k \omega_c t} + \ex^{-\im 2k \omega_c t}\right)\right)\sigma_1
    +\sum_{k=0}^{\infty}\J_{2k+1}(\kappa^\prime_z)\left(\ex^{\im (2k+1) \omega_c t} - \ex^{-\im (2k+1) \omega_c t}\right)\sigma_2,\\
    &U_4\sigma_xU^\dagger_4 = \left(\sigma^+\ex^{\im\omega_c t} + \sigma^-\ex^{-\im\omega_c t}\right),\\
    &U_4\im\sigma_yU^\dagger_4 = \left(\sigma^+\ex^{\im\omega_c t} - \sigma^-\ex^{-\im\omega_c t}\right).
\end{split}
\end{equation*}
where the third line above applies for pairs $(\sigma_1, \sigma_2) = (\sigma_x, \im\sigma_y),\;(\im\sigma_y, \sigma_x)$.

Putting together all the pieces, we end up with a fully transformed Hamiltonian describing system-reservoir interaction
\begin{equation}
    \begin{split}
    H_{R,\text{int}}/\hbar &= \frac{1}{2}\sum_p g_p \underline{\left(b^\dagger_p\ex^{\im\omega_p t} + b_p\ex^{-\im\omega_p t}\right)}\times\\
    \Bigg[
    &-\sum_{k=0}^{\infty}\sum_{l=0}^{\infty}\J_{2k+1}(\kappa_z)\J_{2l+1}(2\xi\kappa_x)\left(\ex^{\im (2k+1) \omega_c t} - \ex^{-\im (2k+1) \omega_c t}\right)
    \left(\ex^{\im (2l+1) \omega_c t} - \ex^{-\im (2l+1) \omega_c t}\right)\sigma_z\\
    &+\Bigg[
    \left( \J_0(\kappa_z) + \sum_{k=1}^{\infty}\J_{2k}(\kappa_z)\left(\ex^{\im 2k \omega_c t} + \ex^{-\im 2k \omega_c t}\right)\right)\left( \J_0(\kappa^\prime_z) + \sum_{l=1}^{\infty}\J_{2l}(\kappa^\prime_z)\left(\ex^{\im 2l \omega_c t} + \ex^{-\im 2l \omega_c t}\right)\right)\\
    &\qquad+\sum_{k=0}^{\infty}\J_{2k+1}(\kappa_z)\left(\ex^{\im (2k+1) \omega_c t} - \ex^{-\im (2k+1) \omega_c t}\right)\left( \J_0(2\xi\kappa_x) + \sum_{l=1}^{\infty}\J_{2l}(2\xi\kappa_x)\left(\ex^{\im 2l \omega_c t} + \ex^{-\im 2l \omega_c t}\right)\right)\times\\
    &\qquad\quad\sum_{m=0}^{\infty}\J_{2m+1}(\kappa^\prime_z)\left(\ex^{\im (2m+1) \omega_c t} - \ex^{-\im (2m+1) \omega_c t}\right)
    \Bigg]
    \underline{\left(\sigma^+\ex^{\im\omega_c t} + \sigma^-\ex^{-\im\omega_c t}\right)}
    \\
    &+\Bigg[
    \left( \J_0(\kappa_z) + \sum_{k=1}^{\infty}\J_{2k}(\kappa_z)\left(\ex^{\im 2k \omega_c t} + \ex^{-\im 2k \omega_c t}\right)\right)\sum_{l=0}^{\infty}\J_{2l+1}(\kappa^\prime_z)\left(\ex^{\im (2l+1) \omega_c t} - \ex^{-\im (2l+1) \omega_c t}\right)\\
    &\qquad+\sum_{k=0}^{\infty}\J_{2k+1}(\kappa_z)\left(\ex^{\im (2k+1) \omega_c t} - \ex^{-\im (2k+1) \omega_c t}\right)\left( \J_0(2\xi\kappa_x) + \sum_{l=1}^{\infty}\J_{2l}(2\xi\kappa_x)\left(\ex^{\im 2l \omega_c t} + \ex^{-\im 2l \omega_c t}\right)\right)\times\\
    &\qquad\quad\left( \J_0(\kappa^\prime_z) + \sum_{m=1}^{\infty}\J_{2m}(\kappa^\prime_z)\left(\ex^{\im 2m \omega_c t} + \ex^{-\im 2m \omega_c t}\right)\right)
    \Bigg]
    \underline{\left(\sigma^+\ex^{\im\omega_c t} - \sigma^-\ex^{-\im\omega_c t}\right)}
    \Bigg].
    \end{split}
\label{eq:HamiltonianR_int}
\end{equation}
Since we consider a thermal reservoir in the vacuum state weakly coupled to the system, the terms oscillating at $\ex^{\pm\im(\omega_p - \omega_c)}$ are significant, while other rapidly oscillating terms in the above equation can be disregarded.
Moreover, we can also exclude the whole second rectangular bracket, as all the multiplications provide time-dependent terms. Hence, the complexity of the problem reduces significantly and we arrive at approximated Hamiltonian given by Eq.~\eqref{eq:HamiltonianRintApprox} in the main text. The full form of the modified coupling strengths $g_p^\prime$ reads
\begin{equation}
    \begin{split}
    g_p^\prime = g_p &\Big( \J_0(\kappa_z) \J_0(\kappa^\prime_z)
    + 2 \sum_{k=1}^\infty\J_{2k}(\kappa_z)\J_{2k}(\kappa^\prime_z) 
    - 2\J_0(2\xi\kappa_x)\sum_{k=0}^\infty\J_{2k+1}(\kappa_z)\J_{2k+1}(\kappa^\prime_z)\\
    &-2\sum_{k=0}^\infty\sum_{l=1}^\infty \J_{2k+1}(\kappa_z)\J_{2l}(2\xi\kappa_x)\J_{2(k+l)+1}(\kappa^\prime_z) 
    + 2\sum_{k=0}^\infty\sum_{m=0}^\infty\J_{2k+1}(\kappa_z)\J_{2(k+m+1)}(2\xi\kappa_x)\J_{2m+1}(\kappa^\prime_z)\\
    &-2\sum_{l=1}^\infty\sum_{m=0}^\infty\J_{2(l+m)+1}(\kappa_z)\J_{2l}(2\xi\kappa_x)\J_{2m+1}(\kappa^\prime_z)\Big)\approx g_p \J_0(\kappa_z) \J_0(\kappa^\prime_z).
    \end{split}
\label{eq:alpha}
\end{equation}
The approximation proposed at the end contains the only term proportional to $\J_0(\kappa_z^\prime)$. Typically $\kappa_z^\prime \ll \kappa_z$, hence the higher order terms do not provide significant corrections: $\J_n(\kappa_z^\prime)\approx0$, for $n\geq1$. 

\section{Appendix: Stationary solution}
\label{sec:AppendixB}
We can write the master equation for the polar system's density matrix $\rho$, which undergoes the unitary transformations $\rho \rightarrow U_4(t)U_3(t) U_2(t) U_1(t) \rho U^\dagger_1(t) U^\dagger_2(t) U^\dagger_3(t) U^\dagger_4(t)$, under the evolution of the effective Hamiltonian from Eq.~\eqref{eq:HamiltonianEff}, as
\begin{equation}
    \dot{\rho} = [H_\text{eff}/\hbar, \rho] - \im\frac{{\gamma^\prime}}{2}\left(\sigma^+\sigma^-\rho + \rho \sigma^+\sigma^- - 2\sigma^-\rho\sigma^+ \right),
\label{eq:masterEQ}
\end{equation}
where the effective rate $\gamma^\prime$ is given by Eq.~\eqref{eq:gamma_prime}.

To solve this equation for the stationary state ($\dot{\rho}(t)=0$), we rewrite it in the matrix form for the expectation values of the operators $\sigma^+$, $\sigma-$, and $\sigma_z$
\begin{equation}
    \vec{0}=
    \begin{pmatrix}
    -\im{\delta^\prime}-\frac{1}{2}{\gamma^\prime} & 0 & -\frac{1}{2}\im{\Omega^\prime}\\
    0 & \im{\delta^\prime} - \frac{1}{2}{\gamma^\prime} & \frac{1}{2}\im{\Omega^\prime}\\
    -\im{\Omega^\prime} & \im{\Omega^\prime} & -{\gamma^\prime}
    \end{pmatrix}
    \begin{pmatrix}
    \langle\sigma^+\rangle_s\\
    \langle\sigma^-\rangle_s\\
    \langle\sigma_z\rangle_s
    \end{pmatrix}
    -\begin{pmatrix}
    0\\
    0\\
    {\gamma^\prime}
    \end{pmatrix},
\label{eq:stationarymatrix}
\end{equation}
where subscript \textit{s} indicates stationary solution. Straightforward algebraic transformations provide expressions for $\langle\sigma_z\rangle_s$ and $\langle\sigma^\pm\rangle_s$ given by Eqs.~\eqref{eq:stationarysolutions}.

\section{Appendix: Fluorescent spectra}
\label{sec:AppendixC}
The general formula for the emission spectrum reads \cite{scully}
\begin{equation}
    S(\omega) = \frac{1}{2\pi}\lim\limits_{T \to \infty}\frac{1}{T}\int_0^T dt \int_0^T dt^\prime g^{(1)}(t, t^\prime)\ex^{-\im\omega(t-t^\prime)}, 
\label{eq:spectrum}
\end{equation}
where $g^{(1)}$ is the first-order correlation function
\begin{equation}
    g^{(1)}(t, t^\prime) = \langle U^\dagger (t)\sigma^+U(t)U^\dagger (t^\prime)\sigma^-U(t^\prime)\rangle,
\label{eq:g1_general}
\end{equation}
where $U(t)$ is an evolution operator of Hamiltonian Eq.~\eqref{eq:full_Hamiltonian}. The further calculations in this section provide generalization of the work done in Ref.~\cite{yan2013}. The correlation function can be found analytically in the reference frame rotated according to the sequence of operators $U_1$--$U_4$
\begin{equation*}
\begin{split}
    &\sigma^\pm \rightarrow U_4(t)U_3(t) U_2(t) U_1(t) \sigma^\pm U^\dagger_1(t) U^\dagger_2(t) U^\dagger_3(t) U^\dagger_4(t)\\
    &= \frac{1}{2}{\sum_{m=0}^\infty \Big[\left(
    \alpha_{m}^z \ex^{\pm\im m \omega_c t} + \beta_m^z\ex^{\mp\im m\omega_c t}
    \right)}\sigma_z + \left(\alpha_m\ex^{\pm\im m\omega_c t} + \beta_m\ex^{\mp\im m\omega_c t} \right)\sigma^\pm + \left(\alpha^\prime_m\ex^{\pm\im m\omega_c t} + \beta^\prime_m\ex^{\mp\im m\omega_c t} \right)\sigma^\mp
    \Big],
\label{eq:sigmapm_unitary}
\end{split}
\end{equation*}
where the $\sigma^\pm$ operators on the right-hand side are given in the rotated reference frame. Above, we have used the following symbols
\begin{equation}
\begin{split}
    &\alpha_m=
    \begin{cases}
    \frac{1}{2}\sum\limits_{n\text{ odd}}^\infty\J_n(\kappa_z^\Sigma)(j_{n+2}-j_n), & m = 0,\\
    \J_0(\kappa_z^\Sigma)j_1+\sum\limits_{n\text{ odd}}^\infty\big[\J_{n+1}(\kappa_z^\Sigma)j_{n+2}
    +\J_{\underset{(n>1)}{n-1}}(\kappa_z^\Sigma)j_n \big], & m = 1,\\
    \sum\limits_{n\text{ odd}}^\infty\big[ \J_{\underset{(m>n)}{m-n}}(\kappa_z^\Sigma)j_n+\J_{m+n}(\kappa_z^\Sigma)j_{n+2}
    -\J_{\underset{(n>m)}{n-m}}(\kappa_z^\Sigma)j_n\big], & m\text{ even},\\
     \sum\limits_{n\text{ odd}}^\infty\big[ \J_{\underset{(m>n)}{m-n}}(\kappa_z^\Sigma)j_n+\J_{m+n}(\kappa_z^\Sigma)j_{n+2}
    +\J_{\underset{(n>m)}{n-m}}(\kappa_z^\Sigma)j_n\big] + J_0(\kappa_z^\Sigma)j_m, & m\text{ odd},
    \end{cases}
\end{split}
\label{eq:alphas}
\end{equation}
\begin{equation}
\begin{split}
    &\beta_m=
    \begin{cases}
    \frac{1}{2}\sum\limits_{n\text{ odd}}^\infty\J_n(\kappa_z^\Sigma)(j_{n+2}-j_n), & m = 0,\\
    \J_0(\kappa_z^\Sigma)j_{1+2}+\sum\limits_{n\text{ odd}}^\infty\big[\J_{n+1}(\kappa_z^\Sigma)j_{n}
    +\J_{\underset{(n>1)}{n-1}}(\kappa_z^\Sigma)j_{n+2}\big], & m = 1,\\
    -\sum\limits_{n\text{ odd}}^\infty\big[ \J_{\underset{(m>n)}{m-n}}(\kappa_z^\Sigma)j_{n+2}+\J_{m+n}(\kappa_z^\Sigma)j_{n}
    -\J_{\underset{(n>m)}{n-m}}(\kappa_z^\Sigma)j_{n+2}\big], & m\text{ even},\\
     \sum\limits_{n\text{ odd}}^\infty\big[ \J_{\underset{(m>n)}{m-n}}(\kappa_z^\Sigma)j_{n+2}+\J_{m+n}(\kappa_z^\Sigma)j_n
     +\J_{\underset{(n>m)}{n-m}}(\kappa_z^\Sigma)j_{n+2}\big] + J_0(\kappa_z^\Sigma)j_{m+2}, & m\text{ odd},
    \end{cases}
\end{split}
\label{eq:betas}
\end{equation}
where
\begin{equation}
\begin{split}
    j_n&= 
    \begin{cases}
    1+\J_0(2\xi\kappa_x), & n = 1,\\
    \J_{n-1}(2\xi\kappa_x), & n \neq 1,
    \end{cases}\\
    j^\prime_n&=
    \begin{cases}
    1-\J_0(2\xi\kappa_x), & n = 1,\\
    -\J_{n-1}(2\xi\kappa_x), & n \neq 1.    
    \end{cases}
\end{split}
\label{eq:jns}
\end{equation}
The relations between the $z$, non-primed, and primed parameters are
\begin{equation}
    \alpha_m^z, \; \beta_m^z \xleftarrow[\substack{j^\prime_{n+1}\leftarrow j_n\\j_{n+1}\leftarrow j_{n+2}\\\kappa_z\leftarrow \kappa_z^\Sigma}]{}\; \alpha_m,\;\beta_m \xrightarrow[\substack{j_n\rightarrow j^\prime_{n+2}\\j_{n+2}\rightarrow j^\prime_n\\\kappa_z^\Sigma\rightarrow \kappa_z^\Delta}]{}\alpha^\prime_m,\;\beta^\prime_m.
\label{eq:alpha_beta_prim}
\end{equation}
Above, $\kappa_z^\Sigma = \kappa_z + \kappa^\prime_z$, and $\kappa_z^\Delta = \kappa_z - \kappa^\prime_z$. Eventually, the correlation function reads as
\begin{equation}
\begin{split}
    g^{(1)}(t, t^\prime) = \Bigg\langle &\frac{1}{2}\sum_{m^\prime=0}^\infty \Big[ \left( 
    \alpha^z_{m^\prime}\ex^{\im m^\prime \omega_c t} + \beta^z_{m^\prime}\ex^{-\im m^\prime\omega_c t}
    \right)\sigma_z\\
    &+\left(\alpha_{m^\prime}\ex^{\im m^\prime\omega_c t} + \beta_{m^\prime}\ex^{-\im m^\prime\omega_c t} \right)\sigma^{+}
    +\left(\alpha^\prime_{m^\prime}\ex^{\im m^\prime\omega_c t} + \beta^\prime_{m^\prime}\ex^{-\im m^\prime\omega_c t} \right)\sigma^{-}
    \Big]\\
    \times&\frac{1}{2}\sum_{m=0}^\infty \Big[\left( 
    \alpha^z_{m}\ex^{-\im m \omega_c t^\prime} + \beta^z_{m}\ex^{\im m\omega_c t^\prime}
    \right)\sigma_z\\
    &+\frac{1}{2}
    \left(\alpha_{m}\ex^{-\im m\omega_c t^\prime} + \beta_{m}\ex^{\im m\omega_c t^\prime} \right)\sigma^{-}
    +\left(\alpha^\prime_{m}\ex^{-\im m\omega_c t^\prime} + \beta^\prime_{m}\ex^{\im m\omega_c t^\prime} \right)\sigma^{+} 
    \Big] \Bigg\rangle.    
    \end{split}
\label{eq:g1}
\end{equation}
We have obtained an expression with two types of terms with exponents: $\ex^{\pm\im m^\prime \omega_c t}\ex^{\mp\im m \omega_c t^\prime}$, and $\ex^{\pm\im m^\prime \omega_c t}\ex^{\pm\im m \omega_c t^\prime}$. To calculate the spectrum based on Eq.~\eqref{eq:spectrum}, we integrate each of these terms, approximating as follows
\begin{equation}
\begin{split}
    I(\omega) &= \lim_{T\rightarrow\infty}\frac{1}{T}\int\limits_{0}^T dt\int\limits_{0}^T dt^\prime\ex^{\pm\im m^\prime \omega_c t}\ex^{\mp\im m \omega_c t^\prime}\ex^{-\im\omega(t-t^\prime)}    \times\langle\sigma_\mu(t-t^\prime)\sigma_\nu(0)\rangle\\
    &=\lim_{T\rightarrow\infty}\frac{1}{T}\int\limits_0^{T}dt^\prime\ex^{\pm\im(m^\prime-m)\omega_c t^\prime}\int\limits_{-t^\prime}^{T-t^\prime}d\tau\ex^{\pm\im(m^\prime\omega_c \mp \omega)\tau}
    \times\langle\sigma_\mu(\tau)\sigma_\nu(0)\rangle\\
    &\approx\delta_{m^\prime,m}\int\limits_{-\infty}^{\infty}d\tau \ex^{\pm\im(m^\prime\omega_c \mp \omega)\tau} \langle\sigma_\mu(\tau)\sigma_\nu(0)\rangle_s,
\end{split}
\label{eq:intensity}
\end{equation}
where $\tau = t-t^\prime$, $\mu, \nu \in \{z, +, - \}$ and we moved to the stationary state for the expectation values $\langle\sigma_\mu(\tau)\sigma_\nu(0)\rangle \rightarrow \langle\sigma_\mu(\tau)\sigma_\nu(0)\rangle_s$ by setting integral limits $T-t^\prime \rightarrow \infty$, $-t^\prime\rightarrow-\infty$. Due to this approximation, we simplified the correlation function by removing one of the sums, as only the terms with $m=m^\prime$ contribute. Moreover, following the same procedure for terms $\ex^{\pm\im m^\prime \omega_c t}\ex^{\pm\im m \omega_c t^\prime}$, we end up with $\delta_{m^\prime, -m}$ which leaves only one term with $m^\prime = -m = 0$, as $m, m^\prime \geq 0$.

We end up with a compact form of the expression for the fluorescence spectrum
\begin{equation}
    S(\omega) = \frac{1}{4\pi} \Re\int\limits_{0}^{\infty}d\tau\sum\limits_{\mu, \nu}\left[A^0_{\mu\nu} + \sum\limits_{n=1}^{\infty}\left(A_{\mu\nu}^{+}(n)\ex^{\im n\omega_c \tau} + A_{\mu\nu}^{-}(n)\ex^{-\im n\omega_c \tau}\right)\right]
    \ex^{-\im \omega \tau}\langle\sigma_\mu(\tau)\sigma_\nu(0)\rangle_s,
\label{eq:spectrum_full}
\end{equation}
where
\begin{equation}
    A_{\mu\nu}^0 =
    \begin{cases}
        \left(\alpha_0^z\right)^2 + \left(\beta_0^z\right)^2 + 2\alpha_0^z\beta_0^z, & \mu\nu=zz,\\
        \left(\alpha_0\right)^2 + \left(\beta_0\right)^2 + 2\alpha_0\beta_0, & \mu\nu=+-,\\
        \left(\alpha_0^\prime\right)^2 + \left(\beta_0^\prime\right)^2 + 2\alpha_0^\prime\beta_0^\prime, & \mu\nu=-+,\\
        \alpha_0^z\alpha_0 + \beta_0^z\beta_0 + \alpha_0^z\beta_0 + \beta_0^z\alpha_0, & \mu\nu=z-,+z,\\
        \alpha_0^z\alpha_0^\prime + \beta_0^z\beta_0^\prime + \alpha_0^z\beta_0^\prime + \beta_0^z\alpha_0^\prime, & \mu\nu=z+,-z,\\
        \alpha_0\alpha_0^\prime + \beta_0\beta_0^\prime + \alpha_0\beta_0^\prime + \beta_0\alpha_0^\prime, & \mu\nu=++,--.
    \end{cases} 
\label{eq:A0_params}
\end{equation}
\begin{equation}
\begin{split}
    &A_{\mu\nu}^+(n)=
    \begin{cases}
        \left(\alpha_n^z\right)^2, & \mu\nu=zz,\\
        \left(\alpha_n\right)^2, & \mu\nu=+-,\\
        \left(\alpha_n^\prime\right)^2, & \mu\nu=-+,\\
        \alpha_n^z \alpha_n, & \mu\nu=z-,+z,\\
        \alpha_n^z \alpha_n^\prime, & \mu\nu=z+,-z,\\
        \alpha_n \alpha_n^\prime, & \mu\nu=++,--.\\
    \end{cases}\\
    &A_{\mu\nu}^+(n) \xrightarrow[\alpha\rightarrow\beta]{}A_{\mu\nu}^-(n).
\end{split}
\label{eq:A_params}
\end{equation}

Finally, we evaluate the expectation values for the operators. We are interested in the incoherent part of the spectrum, hence, we use the quantity
\begin{equation}
    \langle\langle \sigma_\mu(\tau)\sigma_\nu(0) \rangle\rangle = \langle \sigma_\mu(\tau)\sigma_\nu(0) \rangle_s - \langle \sigma_\mu\rangle_s \langle \sigma_\nu\rangle_s.
\label{eq:twotime_correlation}
\end{equation}
Using the quantum regression theorem \cite{scully} and following reasoning in Ref.~\cite{yan2013} for the master equation Eq.~\eqref{eq:master}, we find that
\begin{equation}
    \frac{d }{d\tau} \langle\langle \sigma_\mu(\tau)\sigma_\nu(0) \rangle\rangle = \sum\limits_{\lambda}M_{\mu\lambda} \langle\langle \sigma_\lambda(\tau)\sigma_\nu(0) \rangle\rangle,
\end{equation}
where $M$ is the matrix responsible for the 
homogeneous evolution
\begin{equation}
    M = \begin{pmatrix}
    -\im{\delta^\prime}-\frac{1}{2}{\gamma^\prime} & 0 & -\frac{1}{2}\im{\Omega^\prime}\\
    0 & \im{\delta^\prime} - \frac{1}{2}{\gamma^\prime} & \frac{1}{2}\im{\Omega^\prime}\\
    -\im{\Omega^\prime} & \im{\Omega^\prime} & -{\gamma^\prime}
    \end{pmatrix}.
\end{equation}
Making use of the Laplace transformation $\mathcal{L}(f(t)) = \int^\infty_0 f(t) \ex^{-st} \de t = \mathcal{L}[f](s)$, this set of equations can be written as
\begin{equation}
    \left(M-\frac{1}{s}I\right)\mathcal{L}[X_\nu](s) = X_{0 \nu},
\end{equation}
where $X_{0\nu}$ provides initial conditions
\begin{equation}
    X_{0\nu} = 
    \begin{pmatrix}
    \langle\langle \sigma^{+}(0)\sigma_\nu(0) \rangle\rangle\\
    \langle\langle \sigma^{-}(0)\sigma_\nu(0) \rangle\rangle\\
    \langle\langle \sigma_z(0)\sigma_\nu(0) \rangle\rangle
    \end{pmatrix},
    \quad
    X_{\nu} = 
    \begin{pmatrix}
    \langle\langle \sigma^{+}(\tau)\sigma_\nu(0) \rangle\rangle\\
    \langle\langle \sigma^{-}(\tau)\sigma_\nu(0) \rangle\rangle\\
    \langle\langle \sigma_z(\tau)\sigma_\nu(0) \rangle\rangle
    \end{pmatrix}.
\end{equation}
The solutions can be expressed as \cite{yan2013}
\begin{equation}
    \langle\langle \sigma_\mu(\tau)\sigma_\nu(0) \rangle\rangle = \sum\limits_{l=1}^{3}R_{\mu\nu}^l\ex^{s_l\tau},
\label{eq:twotime_solved}
\end{equation}
where $R^l_{\mu\nu} = \lim\limits_{s \to s_l} (s-s_l) \mathcal{L}[\langle\langle \sigma_\mu(\tau)\sigma_\nu(0) \rangle\rangle](s)$,
for $s_l$ beeing the roots of the determinant
\begin{equation}
    \det\left(M-sI\right) = s^3+2{\gamma^\prime} s^2+({\Omega^\prime}^2+{\delta^\prime}^2+\frac{5}{4}{\gamma^\prime}^2)s + {\gamma^\prime}(\frac{1}{2}{\Omega^\prime}^2+{\delta^\prime}^2+\frac{1}{4}{\gamma^\prime}^2)=0,
\label{eq:f_func}
\end{equation}
which are of the form $s_1 = -\gamma_1$, $s_2 = -\gamma_2 + \im\widetilde{\Omega}$, $s_3 = -\gamma_2 - \im\widetilde{\Omega}$, and $\gamma_{1,2}, \widetilde{\Omega} \geq 0$.

Hence, the spectrum can be written as
\begin{equation}
    S(\omega) = \frac{1}{4\pi} \Re\sum\limits_{\mu, \nu}\sum\limits_{l=1}^{3}R_{\mu\nu}^l\Bigg[A_{\mu\nu}^0\int\limits_0^\infty d\tau \ex^{-\im\omega\tau+s_l\tau}+\sum\limits_{n=1}^{\infty}
    \left(A^{+}_{\mu\nu}(n)\int\limits_{0}^{\infty}d\tau\ex^{-\im(\omega-n\omega_c)\tau + s_l\tau}
    +A^{-}_{\mu\nu}(n)\int\limits_{0}^{\infty}d\tau\ex^{-\im(\omega+n\omega_c)\tau + s_l\tau}\right)
    \Bigg],
\label{eq:spectrum_incoherent}
\end{equation}
With that, we can calculate the above integrals which describe the lineshapes $L_{n,l}^\pm(\omega)$
\begin{equation}
    L^\pm_{n,l}(\omega) = 
    \begin{cases}
    \dfrac{\gamma_1-\im(\omega \mp n\omega_c)}{\gamma^2_1 + (\omega \mp n\omega_c)^2}, & l=1,\\
    \dfrac{\gamma_2-\im(\omega \mp n\omega_c-\widetilde{\Omega})}{\gamma^2_2 + (\omega \mp n\omega_c - \widetilde{\Omega})^2}, & l=2,\\
    \dfrac{\gamma_2-\im(\omega \mp n\omega_c+\widetilde{\Omega})}{\gamma^2_2 + (\omega \mp n\omega_c + \widetilde{\Omega})^2}, & l=3,
    \end{cases}
\label{eq:lineshapes}
\end{equation}
where for $n=0$ there is no distinction between $\pm$ signs $L^\pm_{0,l}(\omega)\equiv L_{0,l}(\omega)$.
Lineshape intensities $I_{n,l}^\pm$ are given by
\begin{equation}
I^\pm_{n,l} = 
\begin{cases}
    \sum\limits_{\mu,\nu}R_{\mu\nu}^l A^0_{\mu\nu} \equiv I_{0,l}, &n=0,\\
    \sum\limits_{\mu,\nu}R_{\mu\nu}^l A^\pm_{\mu\nu}(n), &n>0.
\end{cases}
\label{eq:n_intensity}
\end{equation}
Finally, the full spectrum can be calculated as
\begin{equation}
    S(\omega) = \frac{1}{4\pi}\sum\limits_{l=1}^3\Bigg[ I_{0,l}L_{0,l}(\omega) + \sum\limits_{n=1}^\infty\left(I^+_{n,l}L^+_{n,l}(\omega) + I^-_{n,l}L^-_{n,l}(\omega)\right)\Bigg].
\label{eq:spectrum_analytical}
\end{equation}
\end{widetext}
\end{appendices}

\bibliography{main}
\end{document}